\documentclass[aps,prl,twocolumn,superscriptaddress]{revtex4-2}
\usepackage{graphicx}
\usepackage{bm}
\usepackage{amsmath,amssymb,amsthm,dsfont,bm}
\usepackage{xcolor}
\usepackage{comment}

\usepackage[squaren]{SIunits}

\newcommand{\avg}[1]{\left\langle #1\right\rangle}

\newcommand{\bra}[1]{\ensuremath{\langle#1|}}
\newcommand{\ket}[1]{\ensuremath{|#1\rangle}}

\newcommand{\nph}{\ensuremath{n_{\text{ph}}}}
\newcommand{\ncell}{\ensuremath{n_{\text{cell}}}}
\newcommand{\Ncell}{\ensuremath{N_{\text{cell}}}}
\newcommand{\he}{$^3\text{He}$ }
\newcommand{\Gsq}{\ensuremath{\Gamma_{\text{sq}}}}
\newcommand{\var}{\ensuremath{\text{Var}}}

\usepackage[applemac]{inputenc}

\begin{document}

\title{Nuclear spin squeezing in Helium-3 by continuous quantum nondemolition measurement}

\author{Alan Serafin}
	\affiliation{Laboratoire Kastler Brossel, ENS-Universit\'e PSL, CNRS, Universit\'e de la Sorbonne et Coll\`ege de France, 24 rue Lhomond, 75231 Paris, France}

\author{Matteo Fadel}
	\affiliation{Department of Physics, University of Basel, Klingelbergstrasse 82, 4056 Basel, Switzerland}

\author{Philipp Treutlein}
	\affiliation{Department of Physics, University of Basel, Klingelbergstrasse 82, 4056 Basel, Switzerland}

\author{Alice Sinatra}
	\affiliation{Laboratoire Kastler Brossel, ENS-Universit\'e PSL, CNRS, Universit\'e de la Sorbonne et Coll\`ege de France, 24 rue Lhomond, 75231 Paris, France}

\date{\today}

\begin{abstract}
We propose a technique to control the macroscopic collective nuclear spin of a Helium-3 vapor in the quantum regime using light. The scheme relies on metastability exchange collisions to mediate interactions between optically accessible metastable states and the ground-state nuclear spin, giving rise to an effective nuclear spin-light quantum nondemolition interaction of the Faraday form. Our technique enables measurement-based quantum control of nuclear spins, such as the preparation of spin-squeezed states. This, combined with the day-long coherence time of nuclear spin states in Helium-3, opens the possibility for a number of applications in quantum technology.
\end{abstract}

\maketitle

\textit{Introduction.}
The nuclear spin of Helium-3 atoms in a room-temperature gas is a very well isolated quantum system featuring record-long coherence times of up to several days \cite{He3Review}. It is nowadays used in a variety of applications, such as magnetometry \cite{Heil2017a}, gyroscopes for navigation \cite{Kitching2011a}, as target in particle physics experiments \cite{He3Review}, and even in medicine for magnetic resonance imaging of the human respiratory system \cite{Couch15}. Moreover, Helium-3 gas cells are used for precision measurements in fundamental physics, \textit{e.g.} in the search for anomalous forces \cite{Vasilakis09} or violations of fundamental symmetries in nature \cite{Heil2013a}.

While the exceptional isolation of Helium-3 nuclear spins is key to achieving long coherence times, it renders measurement and control difficult. 
Remarkably, noble gas nuclear spins can be polarized by metastability-exchange or spin-exchange optical pumping, harnessing collisions between atoms in different states or of different species that transfer the optically induced electronic polarisation to the nuclei \cite{He3Review,Batz2011a}. 
However, the role of quantum coherence, quantum noise and many-body quantum correlations in this process is only beginning to be studied \cite{Dantan2005a,Reinaudi2007a,Katz2019a}. 
Optical quantum control of noble gas nuclear spin ensembles is still in an early stage of development, and key concepts of quantum technology such as the generation of non-classical states for quantum metrology \cite{Pezze2018a} or the storage of quantum states of light \cite{Bussieres2013} have not yet been demonstrated.

\begin{figure}
  \centering
\includegraphics[width=0.45\textwidth]{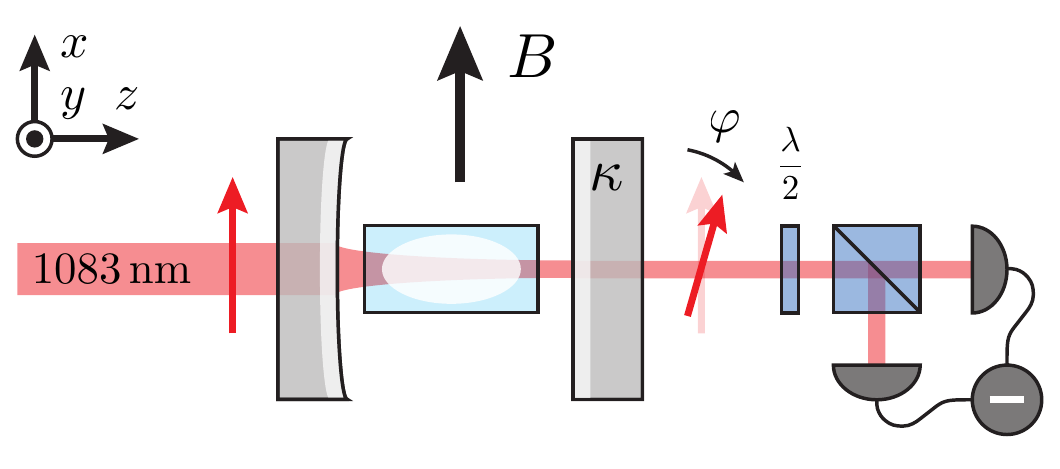}
  \caption{Illustration of the proposed setup. A Helium-3 vapor cell is placed inside an asymmetric optical cavity, ensuring that photons leave the cavity at rate $\kappa$ predominantly through the out-coupling mirror. A (switchable) discharge maintains a small fraction of the atoms in a metastable state. 
  The atomic metastable and nuclear spins are oriented in the $x$ direction beforehand by optical pumping. The light polarization, initially along $x$, is rotated by an angle $\varphi$ due to the Faraday effect, performing a quantum nondemolition measurement of the nuclear spin fluctuations along the light propagation direction. This polarization rotation is continuously monitored via homodyne measurement.}
  \label{figSetup}
\end{figure}

In this paper we propose a technique for the optical manipulation of Helium-3 nuclear spins in the quantum regime. As the nuclear spin state cannot be directly manipulated with light, our approach makes use of metastability exchange collisions to map optically accessible electronic states into the nuclear state, thereby mediating an effective coupling between the light and the nuclear spin. In contrast to earlier ideas put forward by one of us \cite{Dantan2005a,Reinaudi2007a}, the scheme considered here results in a Faraday interaction \cite{HammererRMP10} coupling the fluctuations of the light and of the nuclear spin. This interaction is nowadays routinely used as a powerful and versatile spin-light quantum interface in experiments with alkali vapours \cite{HammererRMP10,StroboNat15}. Since our scheme does not require other atomic species as mediator \cite{Katz2019a,Firstenberg} and the rate constants of metastability exchange collisions are comparatively high \cite{He3Review}, it can operate at room temperature and millibar pressures as commonly used in experiments with Helium-3. Moreover, the interaction can be switched on and off, by switching the week discharge that maintains a population in the metastable state. 
Our scheme will allow to develop quantum-enhanced technologies with Helium-3, such as measurement devices with sensitivity beyond the standard quantum limit \cite{Pezze2018a}.

\begin{figure}
  \centering
\includegraphics[width=0.45\textwidth]{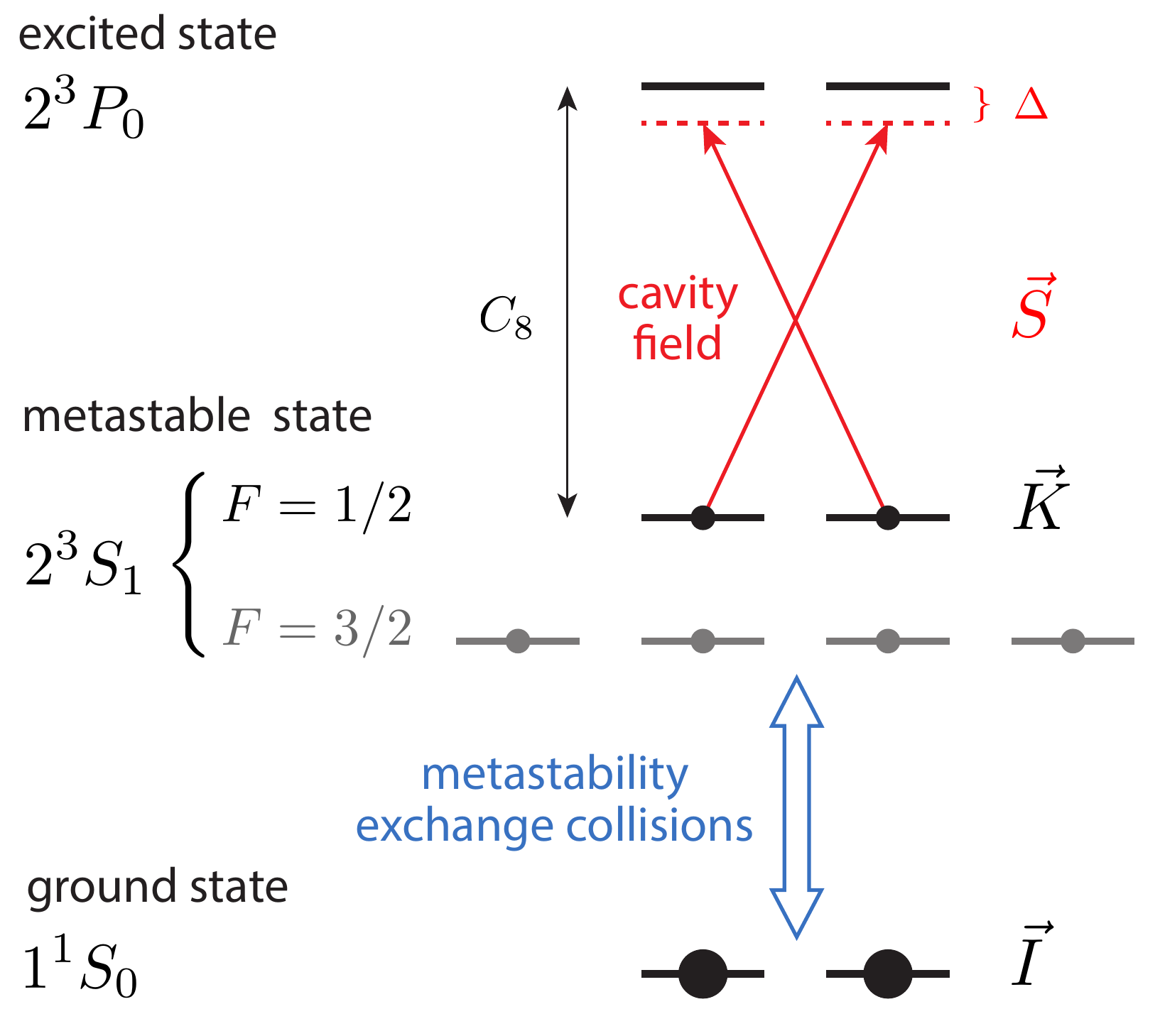}
  \caption{Relevant level scheme of $^3 \text{He}$ for $z$ quantization axis, which corresponds to the cavity axis. The cavity mode (red) addresses the $C_8$ transition between the $F=1/2$ metastable manifold and the $F=1/2$ excited state $2^3 P_0$, with detuning $\Delta$. The six metastable levels $2^3 S_1$ are coupled to the purely nuclear $1^1 S_0$ ground state by metastability exchange collisions.
  }\label{figLev}
\end{figure}

\textit{Semiclassical three-mode model.}
We consider the setup in Fig.~\ref{figSetup}, where a gas cell containing $\Ncell$ Helium-3 atoms in the ground state and a small fraction $\ncell \sim 10^{-6} \Ncell$ in the metastable state is placed inside an optical cavity.
In the theoretical treatment we assume that the metastable atoms are homogeneously illuminated by the cavity mode and the magnetic field is zero. Effects of a small guiding field and the spatial profile of the cavity mode will be discussed at the end of the paper. The relevant level scheme is illustrated in Fig.~\ref{figLev}. We introduce the collective spin operators $\vec{I}$ and $\vec{K}$ for the (nuclear) ground state and for the $F=1/2$ metastable manifold, respectively. For the cavity light, propagating in the $z$-direction and addressing the $2
^{3}S_1-2^{3}P_0$ $C_8$ transition at $\unit{1083}{nm}$, we introduce the Stokes spin operators as a function of the $x$- and $y$-polarized modes as $S_x=(c_x^\dagger c_x - c_y^\dagger c_y)/2$, $S_y=(c_x^\dagger c_y + c_y^\dagger c_x)/2$ and $S_z=(c_x^\dagger c_y - c_y^\dagger c_x)/(2i)$.  
For a large detuning $\Delta$ and in the low-saturation limit, the excited state $2^3 P_0$ can be adiabatically eliminated, resulting in the Faraday interaction Hamiltonian \cite{HammererRMP10}
\begin{equation}
    H = \hbar \chi K_z S_z \label{eq:Faraday_meta}
\end{equation}
with coupling strength $\chi=g_c^2/\Delta$. Here, $g_c = d_8 \mathcal{E}_c / \hbar$ and $\mathcal{E}_c=\sqrt{\frac{\hbar \omega}{2 \epsilon_0 V_c}}$, where $V_c$ is the cavity mode volume, $\omega$ the angular frequency and $d_8$ the dipole matrix element of the chosen transition.

The coupling between $\vec{K}$ and $\vec{I}$ is provided by metastability exchange collisions, occurring at rate $1/\tau$ for a metastable atom, and $1/T$ for a ground state atom, with $T/\tau=\Ncell/\ncell$ \cite{duproc}. 
Metastability exchange collisions can be thought of as an instantaneous exchange of the electronic excitation between a ground state and a metastable atom that leaves nuclear and electronic spins individually unchanged. They are routinely used to transfer orientation between the metastable and the nuclear spins and, as it was shown theoretically, they can also transfer quantum correlations \cite{Dantan2005a,Reinaudi2007a}.
Starting from metastabiliy exchange equations for the metastable and nuclear variables \cite{duproc} plus the Faraday interaction (\ref{eq:Faraday_meta}) between $\vec{K}$ and $\vec{S}$, we write a set of nonlinear equations for the mean values of the collective operators that describe the system dynamics in the semiclassical approximation, i.e. neglecting quantum fluctuations and correlations. 
For $x$-polarized nuclear and light spins 
\begin{equation}
\avg{I_x}_s = {\cal P} \dfrac{\Ncell}{2} \equiv \dfrac{N}{2} \qquad\text{and}\qquad \avg{S_x}_s = \dfrac{\nph}{2} \;,
\end{equation}
where ${\cal P}\in[0,1]$ is the nuclear polarisation and $\nph$ the number of photons in the $c_x$ cavity mode in steady state without atoms, the nonlinear equations of motion admit a stationary solution. In particular, we find
\begin{equation}
\avg{K_x}_s = {\cal P} \left( \dfrac{1-{\cal P}^2}{3+{\cal P}^2} \right) \dfrac{\ncell}{2} \equiv \dfrac{n}{2} \;.
\end{equation}
The nonlinear equations of motion can now be linearized around this stationary solution by setting $\avg{A}=\avg{A}_s+\delta A$, with $A$ a collective operator and $\delta A$ a classical fluctuation. By performing an adiabatic elimination of the $F=3/2$ metastable manifold, we obtain the reduced set of coupled differential equations for the classical fluctuations of the transverse components of three spins
\begin{subequations}\label{eq:SCsystem}
\begin{align}
\dot{\delta S_z} &= -\dfrac{\kappa}{2} \delta S_z \\
\dot{\delta S_y} &= -\dfrac{\kappa}{2} \delta S_y + \chi \avg{S_x}_s \delta K_z \\
\dot{\delta I_z} &= -\gamma_f \delta I_z + \gamma_m \delta K_z \\
\dot{\delta I_y} &= -\gamma_f \delta I_y + \gamma_m \delta K_y \\
\dot{\delta K_z} &= -\gamma_m \delta K_z + \gamma_f \delta I_z \\
\dot{\delta K_y} &= -\gamma_m \delta K_y + \gamma_f \delta I_y + \chi \avg{K_x}_s \delta S_z \;.
\end{align}
\end{subequations}
Here, decay rate and the effective metastability exchange rates for the ground state and metastable atoms are $\gamma_f=\left(\frac{4+{\cal P}^2}{8-{\cal P}^2}\right)\left(\frac{1-{\cal P}^2}{3+{\cal P}^2}\right)\frac{1}{T}$ and $\gamma_m=\left(\frac{4+{\cal P}^2}{8-{\cal P}^2}\right)\frac{1}{\tau}$, respectively. Note that $\gamma_m/\gamma_f = N/n \gg 1$.

We proceed now with a full quantum treatment of the reduced system of three collective spins.

\textit{Quantum three-mode model.} Since $\vec{S}$, $\vec{K}$ and $\vec{I}$ are $x$-polarized and will maintain a large polarization throughout the entire protocol, we can perform the Holstein-Primakoff approximation by replacing $I_y/\sqrt{N}\simeq X_a$, $I_z/\sqrt{N}\simeq P_a$, $K_y/\sqrt{n}\simeq X_b$, $K_z/\sqrt{n}\simeq P_b$,  $S_y/\sqrt{\nph}\simeq X_c$, and $S_z/\sqrt{\nph}\simeq P_c$ where we have introduced the bosonic quadratures $X_\nu=(\nu+\nu^\dagger)/2$, $P_\nu=(\nu-\nu^\dagger)/(2i)$, $[X_\nu,P_\nu]=i/2$ for $\nu=a,b,c$, that describe the transverse fluctuations of the collective spins.
Note that within the Primakoff approximation the mode $c\simeq c_y$
is associated to the $y$-polarized photons inside the cavity. The Faraday Hamiltonian \eqref{eq:Faraday_meta} becomes 
\begin{equation}\label{eq:HamHP}
H = \hbar \Omega P_b P_c \;,
\end{equation}
with $\Omega = \chi \sqrt{n \nph}$. 
In a fully quantum treatment \cite{Dantan2005a}, one adds appropriate Langevin forces representing quantum noise to the semiclassical equations \eqref{eq:SCsystem}. To this approach however, we prefer here an equivalent formulation in terms of a quantum master equation (QME) for the density operator $\rho$ describing the three bosonic modes $a$ (nuclear), $b$ (metastable) and $c$ (cavity), 
\begin{equation}
\dot{\rho} = \frac{1}{i\hbar} [H,\rho] + \sum_{w=c,m} C_w\rho C_w^\dagger - \frac{1}{2}\{C_w^\dagger C_w,\rho\} \;. \label{eq:QME} 
\end{equation}
Besides the interaction Hamiltonian Eq.~\eqref{eq:HamHP}, it includes jump operators for the cavity losses $C_c=\sqrt{\kappa}c$ and for metastability exchange collisions $C_m=-\sqrt{2\gamma_m}b+\sqrt{2\gamma_f}a$. Initially, the three modes are in the vacuum state. Due to the Faraday effect caused by quantum fluctuations of the spin, the polarization of the light is slightly turned and,
after a transient time of order $1/\kappa$, the number of $y$-polarized photons in the cavity reaches the steady state 
\begin{equation}\label{bdaggerb}
    \avg{c^\dagger c}(t) \rightarrow \left(\dfrac{\Omega}{2\kappa}\right)^2 \left( 1 - \dfrac{2 \gamma_m}{\kappa+2(\gamma_m+\gamma_f)}\right) \;.
\end{equation}

\begin{figure*}
  \centering
\includegraphics[width=0.5\textwidth]{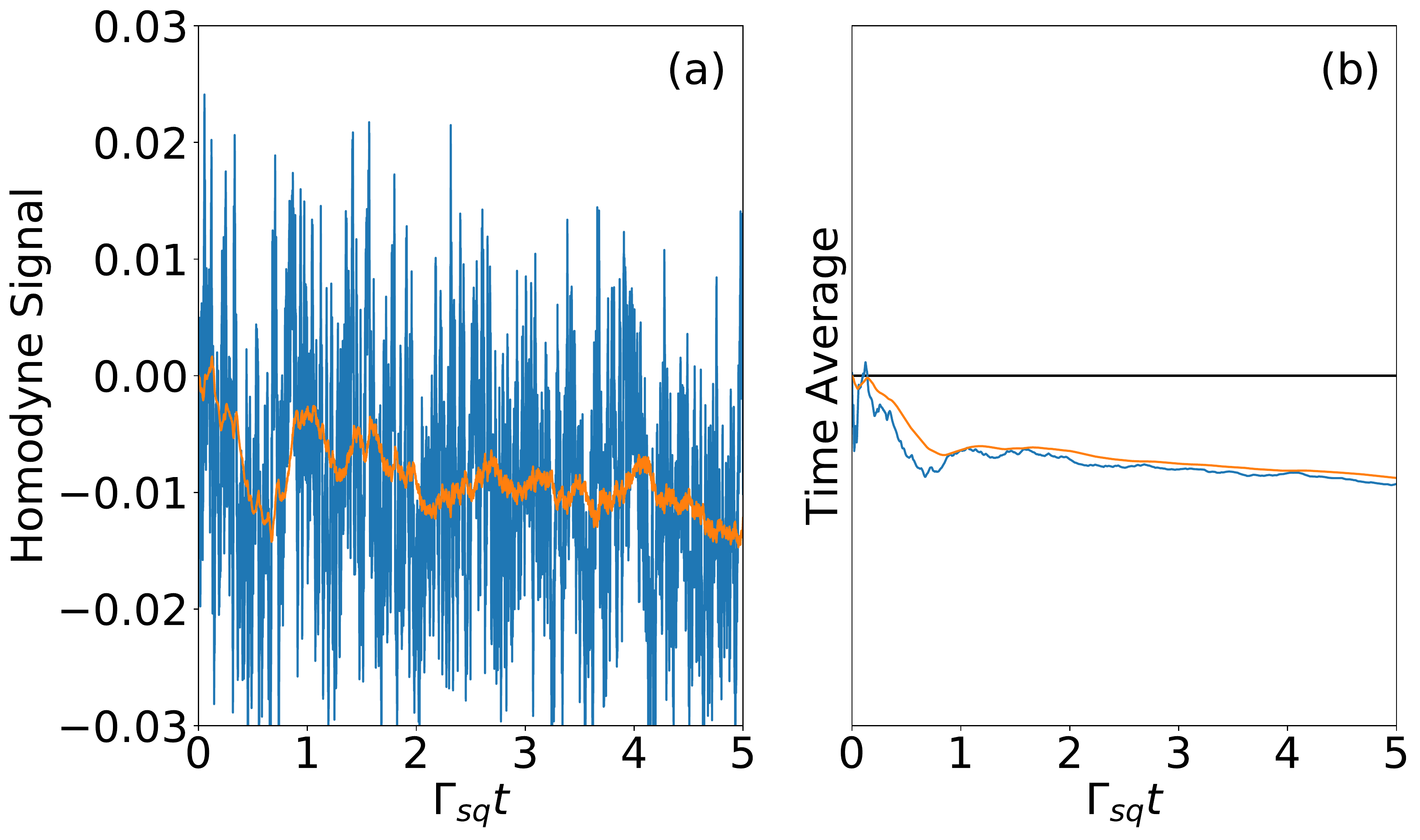}\includegraphics[width=0.45\textwidth]{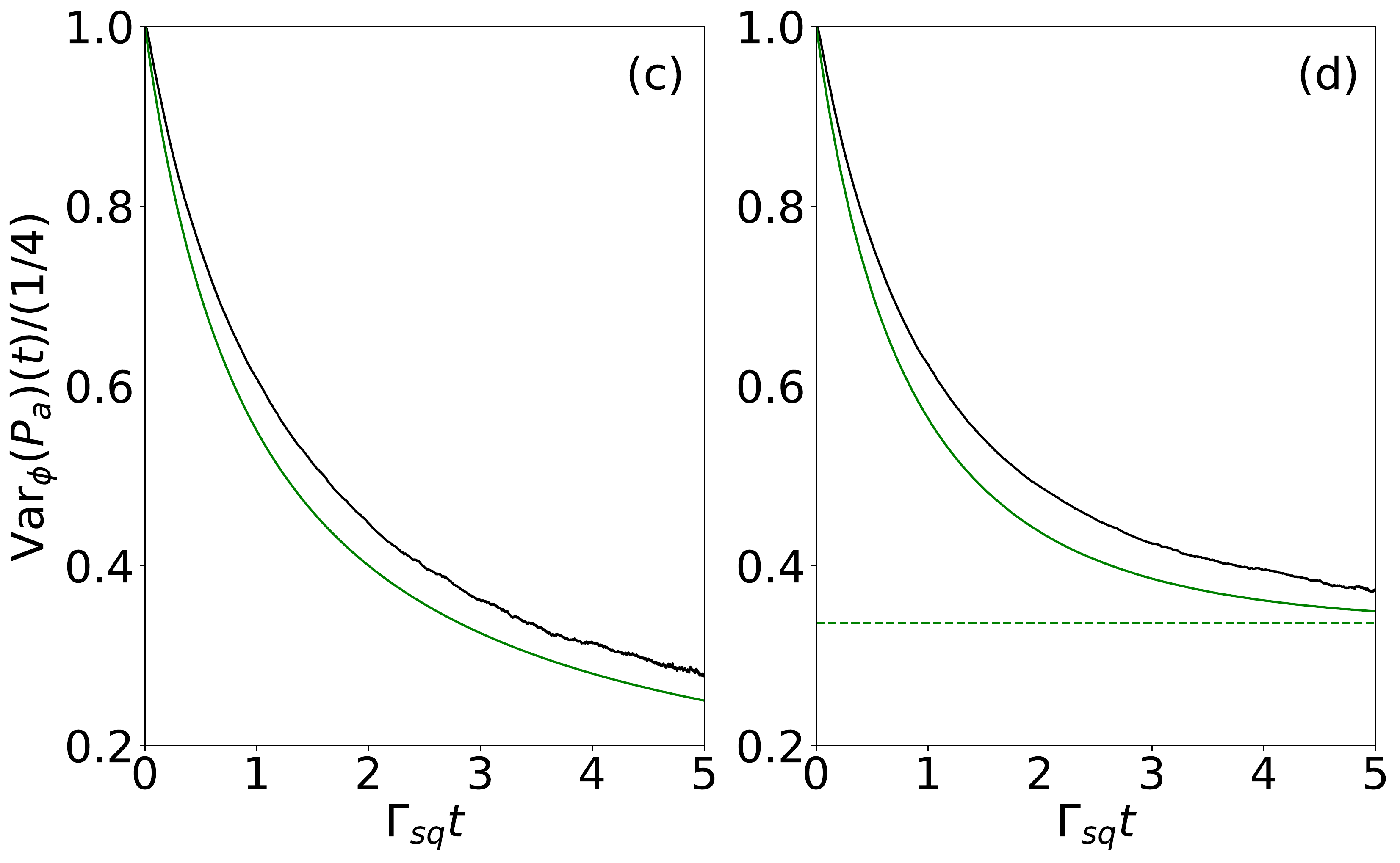} 
  \caption{(a) Time evolution of the homodyne signal $\avg{c+c^\dagger}_\phi$ (blue)
  and of the nuclear spin quadrature $2\sqrt{\frac{\Gamma_{\rm sq}}{\kappa}} \avg{P_a}_\phi$ (orange) in a single realization of the experiment where a continuous homodyne measurement of the $y$-polarized field leaking out of the cavity is performed. (b) Time average of the same quantities. The curves are obtained from the continuous stochastic equation derived from the three-mode QME (\ref{eq:QME}), for a single realization of the stochastic noise describing homodyne detection (the equivalent of $d\zeta_s$ of the one-mode model) and averaged over 5 realization of the stochastic noise describing metastability exchange. Parameters: $\Omega/\kappa = 1/10$, $\gamma_m/\kappa = 1/10$, $\gamma_f/\kappa = 1/100$, $\Gamma_{\rm sq}/\kappa=1/1000$. (c) Conditional variance of the nuclear spin quadrature $P_a$ as a function of time. Black: three-mode model with same parameters as (a), Green: analytical prediction (\ref{varPa}) of the one-mode model. (d) Effect of decoherence. Black: three-mode model with an additional relaxation rate $\gamma_0/\kappa=1/1000$ in the metastable state, where we now average over 8 realizations of the stochastic noises related to metastability exchange and wall relaxation in the metastable state. Green : one-mode model with the corresponding effective relaxation in the ground state $\gamma_0'=\Gamma_{\rm sq}/10$. Dashed horizontal line: analytical prediction (\ref{eq:sqlimit_decoh}).
  }\label{figSim}
\end{figure*}

The metastability exchange collisions lead to a hybridization of the nuclear spin and metastable modes. Their contribution to the three-mode QME is diagonalised introducing the rotated basis
\begin{align}
\label{eq:alpha}
\alpha &= \sqrt{\dfrac{\gamma_m}{\gamma_m+\gamma_f}} a + \sqrt{\dfrac{\gamma_f}{\gamma_m+\gamma_f}} b \;, \\
\label{eq:beta}
\beta &= \sqrt{\dfrac{\gamma_m}{\gamma_m+\gamma_f}} b - \sqrt{\dfrac{\gamma_f}{\gamma_m+\gamma_f}} a \;.
\end{align}
In practice, as $\gamma_m \gg \gamma_f$, $\alpha \approx a$ and $\beta \approx b$. In the rotated basis, the system can be reduced to a one-mode model.

\textit{Reduction to a one-mode model.}
We consider the regime $\kappa \gg \gamma_m \gg \gamma_f$, all being larger than the timescale of the nuclear spin evolution. 
 During the evolution, the number of excitations in the ``hybridized nuclear'' mode $\alpha$ grows linearly in time, while the ``hybridyzed metastable'' mode $\beta$ as well as the cavity mode $c$ will rapidly tend to a stationary value, allowing their adiabatic elimination. 

Following a similar procedure as in Ref.~\cite{CastinPRL} within the Monte-Carlo wavefunction description, we obtain to leading order in the coupling $\Omega$ a one-mode QME describing the slow evolution of the hybridized nuclear mode $\alpha$
\begin{equation}\label{eq:oneQME}
    \dot{\rho}_\alpha = \sum_{w=s,d} \left( C_w\rho C_w^\dagger - \dfrac{1}{2}[C_w^\dagger C_w,\rho] \right) \;.
\end{equation}
This QME involves two jump operators, 
$C_d=\sqrt{\Omega^2/4\kappa}\,\mathbb{I}$ with $\mathbb{I}$ the identity, and $C_s=\sqrt{\Gsq}P_\alpha$ with
\begin{equation}\label{eq:Gsq}
    \Gsq = \dfrac{\Omega^2}{\kappa} \dfrac{\gamma_f}{\gamma_m} \;.
\end{equation}
It appears from the adiabatic elimination that $C_d$ is related to ``double jumps" where a photon and a metastable excitation are annihilated at the same time. This process does not affect the nuclear state vector and it does not play any role in the homodyne-measurement squeezing scheme we consider \footnote{This is because the produced photon is in this case incoherent with the pump and does not contibute to the homodyne signal. It would on the contrary play a role in a scheme based on photon counting as in \cite{WisemanPC}.}. On the contrary we will see that $C_s$, related to single cavity jumps, is responsible for the generation of nuclear spin squeezing at rate $\Gsq$.
Eqs.~(\ref{eq:oneQME},\ref{eq:Gsq}) are one of the main results of our work. The factor $\gamma_f/\gamma_m=n/N$ in Eq.~\eqref{eq:Gsq}, absent in the squeezing rates obtained for alkali atoms using Faraday interactions, reflects the fact that we optically address $n$ metastable atoms to manipulate $N$ nuclear spins.

\textit{Quantum non-demolition measurement of the nuclear spin.}
We now study the evolution of the system in a single experimental realisation, conditioned on the result of a continuous homodyne measurement performed on the small $y$-polarized field leaking out of the cavity, the local oscillator phase being chosen to measure $X_c$ \footnote{Being the conjugate quadrature to $P_c$, $X_c$ carries the information about $P_\alpha$ (see Eqs.~(\ref{eq:HamHP}) and (\ref{eq:alpha})-(\ref{eq:beta}))}. This is described at the level of the QME by appropriate jump operators.
A density matrix conditioned on the measurement can be reconstructed in the Monte Carlo wavefunction method by averaging over stochastic realizations with different histories for metastability exchange collisions but same history for the homodyne detection. 
In the limit of a local oscillator with large amplitude, the evolution of the Monte Carlo wavefunction can be approximated by a nonlinear continuous stochastic evolution \cite{CDM,Gisin}. We apply this approach to both the one-mode model and the three-mode model.

In the case of the one-mode model Eq.~\eqref{eq:oneQME}, the corresponding stochastic evolution reads
\begin{equation}\label{eq:stoc1mode}
  d\ket{\phi(t)} = - \dfrac{dt}{2} \Gsq Q^2 \ket{\phi(t)} + \sqrt{\Gsq} d\zeta_s Q \ket{\phi(t)} \;,
\end{equation}
where $Q\equiv P_\alpha - \bra{\phi}P_\alpha \ket{\phi}$ and $d\zeta_s$ is a real Gaussian random noise of zero mean and variance $d t$. The stochastic equation (\ref{eq:stoc1mode}) describes the evolution of the quantum state of the nuclear spin in a single realization of the experiment. The deterministic term proportional to $\Gamma_{\rm sq} d t$ and the random noise proportional to $\sqrt{\Gamma_{\rm sq}}d\zeta_s$ are issued from the jump operator $C_s$ in the original one mode QME (\ref{eq:oneQME}) and are physically associated to the measurement process on the nuclear spin \cite{Helvetica,Percival,ThomsenPRA02}.
For our initial conditions, the time evolution described by Eq.~\eqref{eq:stoc1mode} can be solved analytically. For a single realization $\phi(t)$ of the stochastic evolution, corresponding to a particular history of homodyne detection, we find that for long times the average $\avg{P_\alpha}_\phi \equiv \bra{\phi}P_\alpha \ket{\phi}$ stabilizes to a (random) constant value, and the variance $\var_\phi (P_\alpha)$ tends to zero as $(\Gsq t)^{-1}$. Going back to the original three-mode basis, the single realisation variance of the nuclear spin quadrature $P_a$ corresponding to $I_z$ reads
\begin{equation}\label{varPa}
    \var_\phi (P_a)(t) = \dfrac{1}{4} \dfrac{1+\frac{\gamma_f}{\gamma_m}\Gsq t}{1 +\Gsq t} \;, 
\end{equation}
and the time average of the homodyne signal is proportional to the fixed (random) value of $\avg{P_a}_\phi$ of that realisation
\begin{equation}\label{eq:sighomo}
    \overline{\avg{c+c^\dagger}_\phi} \stackrel{t\rightarrow\infty}{\longrightarrow}
    2 \sqrt{\frac{\Gamma_{\rm sq}}{\kappa}}\avg{P_a}_\phi \;.
\end{equation}
Note that $\var_\phi (P_a)(t)$ tends to $\gamma_f/(4\gamma_m)$ in the $t\rightarrow\infty$ limit, which is the theoretical spin squeezing limit intrinsic to this method that uses the metastable state to mediate the interaction. In Fig.~\ref{figSim}b-c we compare the analytical predictions (\ref{eq:sighomo}) and (\ref{varPa}) with the numerical solution of the three-mode model.

We note that the limit $\gamma_f/\gamma_m \to 0$ of equation \eqref{varPa} coincides with the result that one would obtain from a nuclear spin-light interaction of the quantum nondemolition or Faraday form
\begin{equation}
\label{eq:HamHP_nucl}
H_{\rm eff} = \hbar \Omega \sqrt{\frac{n}{N}} P_a P_c \quad \mbox{or} \quad H_{\rm eff} = \hbar \chi \frac{n}{N} I_z S_z \:.
\end{equation}

\textit{Effect of decoherence.} 
Due to the long coherence time of the nuclear spin, we can ignore its decoherence on the time scale of squeezing generation. On the other hand, decoherence in the metastable state, including spontanous emission and collisions with the cell walls, will affect the performance of the squeezing protocol. 
From analytical calculations we can show that a relaxation with rate $\gamma_0$ in the metastable state appears in the ground state as an effective relaxation with reduced rate $\gamma_0^\prime = \gamma_0 \frac{\gamma_f}{\gamma_m}$. We thus expect the effect of metastable relaxation to become negligible for $\Gsq \gg \gamma_0^\prime$. By inserting this effective relaxation in the one-mode model \eqref{eq:oneQME}, we calculated the squeezing limit in a single realisation in the presence of metastable decoherence for $\gamma_m \gg \gamma_f$ and $\Gsq \gg \gamma_0^\prime$,
\begin{equation}
     \var_\phi (P_a) \stackrel{t \to \infty}{\longrightarrow} \dfrac{1}{4}\sqrt{\dfrac{\gamma_0^\prime}{\Gsq}} \quad \text{and} \quad
    \var_\phi (X_a) \stackrel{t \to \infty}{\longrightarrow} \dfrac{1}{4}\sqrt{\dfrac{\Gsq}{\gamma_0^\prime}} \;.
    \label{eq:sqlimit_decoh}
\end{equation}
This kind of scaling, already found for alkali atoms \cite{Molmer}, is further confirmed by our numerical simulations where we introduce an additional jump operator $\sqrt{\gamma_0}b$ in the three-mode QME (\ref{eq:QME}), see Fig.~\ref{figSim}d.
An extended theoretical treatment will be published in a separate paper \cite{LongPaperTh}.

\textit{Experimental proposal.} We consider a cylindrical vapor cell $\unit{20}{mm}$ long and $\unit{5}{mm}$ in diameter, filled with $N_\text{cell} = 2.5 \times 10^{16}$ \he atoms at a pressure of $p=\unit{2}{Torr}$. For a polarization of ${\cal P}=0.4$  this gives an effective number of ground state atoms $N = 1.0 \times 10^{16}$. We take $\frac{n_\text{cell}}{N_\text{cell}} = 5\times 10^{-6}$, giving an effective number of metastable atoms $n = 1.3 \times 10^{10}$. From the metastability exchange rate coefficient \cite{He3Review}, we determine effective metastability exchange rates $\gamma_m=\unit{5.2 \times 10^6}{s^{-1}}$ and $\gamma_f = \unit{7.0}{s^{-1}}$.
The cell is placed inside an optical cavity to enhance the atom-light interaction \cite{StroboNat15}. For a finesse of $\mathcal{F}= 50$ and a cavity length of $\unit{3}{cm}$, we obtain $\kappa= 2 \pi \, \unit{1.0 \times 10^8}{Hz}$. 
The cavity is laser driven on the $x$-polarization mode so that $\unit{5}{mW}$ of light exit the cavity in this polarization, and we take the light to be detuned by $\Delta=2\pi \,\unit{2.0}{GHz}$ from the $C_8$ transition. This results in $\Omega=2\pi\,\unit{4.1\times 10^6}{Hz}$. In steady state, $\unit{6.5 \times 10^{5}}{s^{-1}}$ $y$-polarized photons leave the cavity, Eq.~\eqref{bdaggerb}.
The nuclear spin squeezing rate is evaluated from Eq.~\eqref{eq:Gsq} to $\Gamma_{\text{sq}} = \unit{1.4}{s^{-1}}$. We have assumed that atomic motion averages over spatial inhomogeneities of the cavity mode, effectively coupling the light homogeneously to all atoms in the cell \footnote{The squeezing time scale $1/\Gamma_{\text{sq}}$ is long compared to the time scale $1/\gamma_0^{\rm wall}$ for atomic motion between cell walls, $\gamma_0^{\rm wall}/\Gamma_{\text{sq}} \sim 10^4$. The atomic motion thus averages over the spatial variations of the cavity mode, ensuring the validity of a description in terms of collective interactions.}.
From the diffusion coefficient of metastable atoms \cite{Fitz}, we estimate the metastable relaxation rate due to wall collisions to be $\gamma_0^{\rm wall}=\unit{2.6 \times 10^4}{s^{-1}}$ \cite{Franzen}. The off-resonant photon scattering rate in the metastable state, averaged over the cell, is $\gamma_{0}^{\rm scat} \approx \unit{2.4 \times 10^3}{s^{-1}} \ll \gamma_0^{\rm wall}$. 
According to \eqref{eq:sqlimit_decoh}, the squeezing limit for these parameters is $\unit{-8}{dB}$. We note that the squeezing limit imposed by photon scattering is the same as for alkali atoms, since the factor $n/N$ appears both in the effective coupling \eqref{eq:HamHP_nucl} and in the effective nuclear spin decoherence rate $\gamma_0'$ in terms of the metastable decoherence rate $\gamma_0$.
For such squeezing levels, we estimate that the Larmor precession in a small guiding field of $\unit{10^{-7}}{G}$ for $t=\unit{10}{s}$, approximately the whole duration of the experiment, can be neglected 
\footnote{We consider that the effect of a magnetic field $B$ over a time $t$ is negligible if the precession of the noise ellipse of a \unit{10}{dB} squeezed state degrades the squeezed variance by less than 10\% (this corresponds to an angle of 1.8 degrees). Given that the Larmor frequency in the ground state is \unit{3.24}{kHz/G}, we obtain the condition $B[\mathrm{G}]\times t[s] \le 1.5\times 10^{-6}$. Although the Larmor frequency in the metastable state is much larger, \unit{1.87}{MHz/G}, the precession in this state is negligible for magnetic fields up to $\sim \unit{10}{mG}$ since the rotation in the $zy$ plane occurs only during the short time $1/\gamma_m$ between two metastability exchange collisions, corresponding to an angle of order $1$ degree.}.
For larger guiding fields of order \unit{10}{mG}, stroboscopic measurements can be used to evade quantum back-action \cite{StroboNat15}.

\textit{Conclusions.} 
In this work we proposed a technique for the optical manipulation of the \he collective nuclear spin in the quantum regime. In particular, we have shown that QND measurement techniques previously developed for alkali atoms can be generalized to this system, giving access to a measurement-based preparation of nonclassical nuclear spin states, and thus constituting a fundamental building block for Helium-spin based quantum technologies. 
Concrete examples that are realistic for the near future include measurement devices with a sensitivity beyond the classical limit and quantum memories for light with ultra-long (several days) storage times.

\textit{Acknowledgments.} We thank Y. Castin, P.-J. Nacher, G. Tastevin, W. Heil, O. Firstenberg and F. Lalo\"e for the useful discussions. All authors acknowledge funding from the project macQsimal of the EU Quantum Flagship. MF was supported by the Research Fund of the University of Basel for Excellent Junior Researchers.

\end{document}